\documentclass[
letter,
amsfonts, amssymb, amsmath,
preprint,
showkeys,
superscriptaddress
]{revtex4-1}

\usepackage{xcolor}
\usepackage{graphicx} 
\usepackage{subfigure}
\usepackage{dcolumn} 
\usepackage{bm}
\usepackage{float}
\usepackage[columnwise,running]{lineno}
\usepackage[normalem]{ulem}
\usepackage{slashed}
\usepackage{booktabs}
\usepackage{hyperref}
\usepackage{setspace}
\usepackage{color}
\definecolor{orange}{cmyk}{0.,0.353,1.,0.}    % orange

\usepackage{tikz}
\usetikzlibrary{positioning}

\begin{document}
\title{Design Principles for AI-Ready QCD Data with a Barrel Imaging Calorimeter Application}
\author{Zhiwan Xu}
\author{Sylvester Joosten}
\author{Minho Kim}
\author{Chun Yuen Tsang}
\author{Maria \.{Z}urek}
\affiliation{Argonne National Laboratory, 9700 S. Cass Avenue, Lemont, IL 60439, Illinois, U.S.A.}
%\date{\today}

\begin{abstract}
Data from large physics collider experiments in Quantum Chromodynamics (QCD) research differ fundamentally from the modalities used in modern foundation models.
The heterogeneity of detector readouts and their technology dependence require principled curation for cross experiment AI applications.
We present a design framework for AI-ready QCD data to define a unified data structure that accommodates heterogeneous detector technologies within a single schema.
We apply the design principle to the simulated data of the Barrel Imaging Calorimeter (BIC) in the ePIC detector at the Electron--Ion Collider.
The BIC simulation data combines AstroPix silicon pixel imaging layers with Pb/ScFi calorimeter layers across different readout types.
We describe the schema specialization, data preparation pipeline, and visualization of the curated AI-ready dataset.
\end{abstract}
\maketitle
\tableofcontents
\newpage

\section{Introduction}

Nuclear and particle physics detectors record particle interactions as heterogeneous electronic signals. A single collision event produces thousands of readouts across multiple subdetectors, each with distinct geometry, segmentation, readout technology, dynamic range, and noise characteristics. As a result, detector data differ fundamentally from the homogeneous arrays commonly assumed in mainstream LLM and AI workflows.

The heterogeneity makes data representation a central design problem for AI applications in QCD data. Raw detector outputs are often optimized for hardware operation, software reconstruction, or detector-specific analysis, not for large-scale AI research. Without a principled representation, AI applications rely on ad hoc preprocessing pipelines that are difficult to reuse across detector systems, learning tasks, and model architectures.
Meanwhile, existing HEP and NP data formats such as ROOT TTrees, EDM4hep~\cite{EDM4hep}, and HDF5-based representations are designed around physics-motivated data models. They typically employ tree-branch-leaf structures with variable-length containers per event, requiring iterative lookup and looping to access specific information. 
This slows AI training process, which 
%often needs fixed-size array access and fast batch iteration. 
typically relies on uniform, well-structured data layouts that support efficient array-style access and fast batch iteration.
Moreover, these formats do not explicitly align detector measurements with per-measurement truth labels for supervised learning, and operate in a physics-oriented object space rather than the information space expected by AI models.

Recent community efforts have begun to address AI-readiness and FAIR (Findability, Accessibility, Interoperability, Reusability) principles for HEP datasets~\cite{Chen:2021euv}, and benchmark datasets have been developed for specific ML tasks such as jet tagging~\cite{Kansal:2022spb}, tracking~\cite{Amrouche:2019wmx}, and calorimeter simulation. %~\cite{Krause:2024avx}
These efforts, however, target specific physics tasks or detector geometries rather than providing a general-purpose schema for heterogeneous detector-level data.
An AI-ready format must balance several competing goals: preserving physically relevant information, minimizing storage and computational overhead, and providing a consistent structure that can support scalable downstream learning.

This work defines a minimal, technology-agnostic AI-ready data format for QCD detector data. The format accommodates tracking, silicon pixel sensors, calorimeters, and waveform-based readouts within a unified schema. 
The schema should also support streaming readout and real-time data processing, which are key capabilities of the next-generation HEP and NP detectors.
Two arrays compose the core of the format: \textcolor{brown}{\texttt{measurements}}, which encodes detector responses, and \textcolor{brown}{\texttt{labels}}, which records the true physics interactions from the distinct particles that contribute to each response.
The physical interpretation of each field is documented in the dataset metadata, enabling use by non-experts outside the field.

To demonstrate the framework on a concrete detector system, we present a case study of the Barrel Imaging Calorimeter (BIC) in the ePIC experiment at the Electron--Ion Collider (EIC). The BIC is a hybrid electromagnetic calorimeter combining AstroPix silicon pixel imaging layers with Pb/ScFi (Lead/Scintillating Fiber) calorimeter layers. Its readout spans both single-sample pixel measurements and multi-sample waveform signals, making it an ideal system for the heterogeneous representation developed here.

Section~\ref{sec:design} presents the general data design, including the design principles and the definitions of the \textcolor{brown}{\texttt{measurements}} and \textcolor{brown}{\texttt{labels}} arrays. Section~\ref{sec:bic} applies the framework to the BIC, describing the schema specialization, data preparation, and visualization. Section~\ref{sec:summary} summarizes the results and outlook.

\section{Data Design}
\label{sec:design}

\subsection{Design Principles} 
Detector observations are represented through a common spacetime-signal abstraction, \[ (x, y, z, A, t), \] where $(x,y,z,t)$ denote the coordinates associated with a measurement and $A$ its signal amplitude. This representation is detector-agnostic for different technologies, including silicon sensors, calorimeters, and waveform-based systems, which can all be expressed within the same schema once their detector-specific meanings are defined in metadata. The format follows four design principles: 
\begin{enumerate} 
\item \textbf{Unified representation.} A single schema supports all detector systems through the \texttt{detector} index and the \texttt{sample} index of hit samples, covering both single-readout and multi-sample measurements. 
\item \textbf{Explicit uncertainty encoding.} Uncertainty quantifies the statistical confidence associated with the data. Spatial uncertainty is stored as the upper triangle of a $3\times3$ covariance matrix; amplitude and time uncertainties are stored as scalars. 
\item \textbf{Metadata-defined semantics.} Units, coordinate conventions, and detector-specific meanings are specified in metadata rather than embedded in field names. 
\item \textbf{Minimal scope.} The format stores only information needed for AI applications and does not duplicate complete native event records such as HepMC3~\cite{Buckley:2019xhk} or EDM4hep~\cite{EDM4hep}. \end{enumerate}

\subsection{The \textcolor{brown}{\texttt{measurements}} Array} 
The \textcolor{brown}{\texttt{measurements}} array is the primary data structure for detector observations. Each row corresponds to one recorded measurement sample. Depending on the detector technology, a sample may represent a single sensor response or one element of a multi-sample readout such as a waveform. It is designed to support heterogeneous detector systems within a single schema while preserving the information needed for downstream AI applications. Its structure is summarized in Table~\ref{table:measurements}. 

\begin{table}[ht] 
\centering \caption{Structure of the \textcolor{brown}{\texttt{measurements}} array.} \begin{tabular}{lll} 
\toprule 
\hline \hline

Field & Type & Description \\
\midrule 
\hline
\texttt{event} & \texttt{uint64} & Event identifier. \\ \texttt{detector} & \texttt{uint16} & Subdetector identifier. \\ \texttt{hit} & \texttt{uint16} & Hit index within a subdetector. \\ \texttt{sample} & \texttt{uint16} & Sample index within a hit. \\ \texttt{x, y, z} & \texttt{float32}$\times$3 & Spatial coordinates associated with the sample. \\ \texttt{pos\_cov\_xx, xy, xz, yy, yz, zz} & \texttt{float32}$\times$6 & Upper triangle of the position covariance matrix. \\ \texttt{amplitude} & \texttt{float32} & Measured signal amplitude. \\ \texttt{amplitude\_uncertainty} & \texttt{float32} & Uncertainty on the amplitude. \\ \texttt{time} & \texttt{float32} & Time associated with the measurement. \\ \texttt{time\_uncertainty} & \texttt{float32} & Uncertainty on the time measurement. \\ 
\hline \hline

\bottomrule \end{tabular} \label{table:measurements} \end{table} 

\paragraph{Field definitions.} 
\begin{itemize} \item \textbf{\texttt{event}:} Identifier for an event in physics experiment. An event denotes an arbitrary grouping of measurements, such as a physics collision, a time-ordered DAQ readout, or another unit defined in the metadata. 
\item \textbf{\texttt{detector}:} Identifier for the subdetector that produced the measurement. The mapping from identifier to detector name and properties is provided in the metadata. 
\item \textbf{\texttt{hit}:} Index for a distinct measurement instance within a subdetector. For example, this may correspond to a sensor hit, a channel response, or another detector-defined interaction unit. 
\item \textbf{\texttt{sample}:} Sub-index within a hit, providing a unified representation for both single-sample and multi-sample readouts. For single-sample measurements, \texttt{sample} is zero. For waveform-like readouts, each time sample is stored in a separate row with the same \texttt{hit} index. 
\item \textbf{\texttt{x, y, z}:} Spatial coordinates associated with the measurement. These typically correspond to the location of the physical readout element, because the true interaction position may require reconstruction. If clustering or other upstream processing has been applied, they may instead represent an estimated interaction position. Units and coordinate conventions are specified in the metadata. 
\item \textbf{\texttt{pos\_cov\_xx, xy, xz, yy, yz, zz}:} Upper triangle of the $3\times3$ covariance matrix associated with the spatial coordinates. This field encodes measurement resolution and correlations among spatial components. 
\item \textbf{\texttt{amplitude}:} Measured signal size. Depending on the detector, this may represent deposited energy, charge, ADC value, photoelectron count, or another \sout{detector-specific} observable. The interpretation and units are defined in the metadata. 
\item \textbf{\texttt{amplitude\_uncertainty}:} Uncertainty associated with the stored amplitude measurement. It applies to the recorded observable and does not necessarily represent the uncertainty on a derived reconstructed quantity. 
\item \textbf{\texttt{time}:} Time associated with the measurement. This may correspond to the detector readout time or to a value assigned during upstream processing. For multi-sample readouts, per-sample times allow reconstruction of the full waveform. Units and conventions are specified in the metadata. 
\item \textbf{\texttt{time\_uncertainty}:} Uncertainty associated with the stored time measurement, reflecting detector and electronics timing resolution. 
\end{itemize}

Each row in \textcolor{brown}{\texttt{measurements}} represents a single point in a point cloud representation, such as a hit or waveform sample with spatial coordinates and signal features. This is crucial for foundation model application. For graph neural network (GNN) or transformer architectures, these rows can be treated as nodes. Edges can be defined at input time using either geometric or topological criteria, such as 
$k$-nearest neighbors, detector adjacency, or shared hit identifiers. 
These connectivity choices are made during data loading.
The \texttt{detector} field can also be used as a heterogeneous node type for models that learn subsystem specific representations.

\section{Truth Label Design}
\subsection{Design Principles} The labels are truth particle interaction information associated with individual measurements of the detector. Each label is defined at the measurement level so that a recorded detector response (a measurement) can be linked directly to the particle(s) that generated the interaction being measured and are responsible for it. 
This design enables fine-grained detector-level truth labeling for downstream supervised learning tasks while remaining applicable across different detector technologies. 
The label format is guided by five principles: 
\begin{enumerate} \setcounter{enumi}{0}
\item \textbf{Measurement-level granularity.} Truth is stored per measurement rather than only per event or reconstructed object, enabling direct supervised learning at the level of recorded detector response.
\item \textbf{Top-$N$ contributors.} For each measurement, the $N$ contributing MC particles with the largest deposited energy are retained. This captures the dominant physical content of the signal while keeping memory usage bounded.
\item \textbf{Spatiotemporal truth description.} Each measurement label includes truth-level spatial coordinates, time, and total deposited energy, together with the corresponding quantities for each retained contributor.
\item \textbf{Particle decay chain association.} Contributor labels may be traced back to a physically relevant ancestor particle when needed. For example, in a calorimeter this prevents labels from being dominated by low-energy shower secondaries instead of the incident particle.
\item \textbf{Minimal information.} A limited amount of information is retained for each contributor. This is sufficient to capture common production and decay patterns without duplicating the full event-level truth record. 
\end{enumerate}

\subsection{The \textcolor{brown}{\texttt{labels}} Array} 

The \textcolor{brown}{\texttt{labels}} array stores measurement-level truth in a tabular form aligned with the \textcolor{brown}{\texttt{measurements}} array. Each row corresponds to one measurement and contains aggregate truth information together with per-particle contributor truth fields for up to $N$ retained particles, indexed by $n \in \{0,\ldots,N-1\}$. This structure allows a detector response to be linked both to its overall truth quantities and to the dominant particle contributions that generated it. Its structure is summarized in Table~\ref{table:labels}. 

\begin{table}[ht] \centering \caption{Structure of the \textcolor{brown}{\texttt{labels}} array, with contributor-specific fields repeated for each retained contributor $n \in \{0,\ldots,N-1\}$.} 
\begin{tabular}{lll} 
\toprule \hline \hline 
Field & Type & Description \\ \hline 
\midrule 
\texttt{event} & \texttt{uint64} & Event identifier. \\ 
\texttt{detector} & \texttt{uint16} & Subdetector identifier. \\
\texttt{hit} & \texttt{uint16} & Hit index within a subdetector. \\
\texttt{sample} & \texttt{uint16} & Sample index within a hit. \\
\texttt{x, y, z} & \texttt{float32}$\times$3 & Truth spatial coordinates. \\
\texttt{deposit\_energy} & \texttt{float32} & Truth total deposited energy. \\
\texttt{time} & \texttt{float32} & Truth time associated with the measurement. \\
\texttt{n\_contributions} & \texttt{uint16} & Total number of contributing particles. \\
\texttt{con\{n\}\_particle} & \texttt{uint32} & Index of contributor $n$ in the event-level record. \\
\texttt{con\{n\}\_x, y, z} & \texttt{float32}$\times$3 & Interaction spatial coordinate of contributor $n$. \\
\texttt{con\{n\}\_deposit\_energy} & \texttt{float32} & Deposited energy from contributor $n$. \\
\texttt{con\{n\}\_time} & \texttt{float32} & Time of interaction of contributor $n$. \\
\texttt{con\{n\}\_px, py, pz} & \texttt{float32}$\times$3 & Momentum vector of contributor $n$ at production. \\
\texttt{con\{n\}\_energy} & \texttt{float32} & Energy of contributor $n$ at production. \\
\texttt{con\{n\}\_pid} & \texttt{int32} & PDG identifier of contributor $n$. \\
\texttt{con\{n\}\_parent\_pid} & \texttt{int32} & PDG identifier of the immediate parent of contributor $n$. \\
\texttt{con\{n\}\_parent\_energy} & \texttt{float32} & Energy of the immediate parent of contributor $n$. \\ \hline \hline 
\bottomrule 
\end{tabular} \label{table:labels} \end{table} 

\paragraph{Field definitions.} 
\begin{itemize} \item \textbf{\texttt{event}, \texttt{detector}, \texttt{hit}, \texttt{sample}:} These fields mirror the corresponding identifiers in \textcolor{brown}{\texttt{measurements}} and allow row-level association between the two arrays. Depending on the detector readout, a label may correspond to a full hit or to an individual sample within a multi-sample measurement. The precise interpretation is documented in the metadata. 
\item \textbf{\texttt{x, y, z}:} Truth coordinates associated with the measurement. These represent the reference spatial location used for labeling and may correspond to a simulated interaction point, an energy-weighted position, or another detector-specific definition documented in the metadata. 
\item \textbf{\texttt{deposit\_energy}:} Total deposited energy associated with the measurement, summed over all contributing particles. This field is the truth counterpart to the signal amplitude stored in \textcolor{brown}{\texttt{measurements}}, although the exact relationship is detector-dependent and defined in the metadata.
\item \textbf{\texttt{time}:} Truth time associated with the measurement. This typically corresponds to the earliest contributing interaction time, but other definitions may be used  and should be documented in the metadata. 
\item \textbf{\texttt{n\_contributions}:} Total number of distinct particle contributions to the measurement. Full contributor-level information is stored only for the top-$N$ contributors; this field preserves the total multiplicity and encodes the degree of signal overlap. 
\item \textbf{\texttt{con\{n\}\_particle}:} Index linking contributor $n$ to an external event-level particle truth record. The full event-level particle list is not included, since widely used community standards such as HepMC3~\cite{Buckley:2019xhk} already provide it. Event-level truth can be stored separately and joined through this index. 
\item \textbf{\texttt{con\{n\}\_x, y, z}:} Truth interaction coordinates associated with contributor $n$. Their exact definition depends on the labeling procedure and detector response model, and is specified in the metadata. 
\item \textbf{\texttt{con\{n\}\_deposit\_energy}:} Deposited energy from contributor $n$ within the labeled measurement. 
\item \textbf{\texttt{con\{n\}\_time}:} Truth interaction time associated with contributor $n$. 
\item \textbf{\texttt{con\{n\}\_px, py, pz}:} Production-level momentum components of contributor $n$, defined at the particle vertex before its interaction in the detector. 
\item \textbf{\texttt{con\{n\}\_energy}:} Production-level energy of contributor $n$, defined at the particle vertex before its interaction in the detector. 
\item \textbf{\texttt{con\{n\}\_pid}:} PDG particle identifier~\cite{PDG:2022pth} of contributor $n$. 
\item \textbf{\texttt{con\{n\}\_parent\_pid}:} PDG particle identifier of the immediate parent of contributor $n$. This enables limited ancestry tracing, such as identifying photons from $\pi^0 \rightarrow \gamma\gamma$ or leptons from heavy-particle decays. The value is set to zero when parent information is unavailable or not applicable. 
\item \textbf{\texttt{con\{n\}\_parent\_energy}:} Production-level energy of the immediate parent of contributor $n$. The value is set to zero when parent information is unavailable or not applicable. 
\end{itemize}

\section{Application to BIC}
\label{sec:bic}

The Barrel Imaging Calorimeter (BIC) is a detector subsystem of the ePIC detector at the Electron--Ion Collider (EIC)~\cite{AbdulKhalek:2021gbh}.  The BIC is a hybrid electromagnetic calorimeter with $2\pi$ azimuthal coverage, inner radius of $r \approx 0.815$~m, and longitudinal extent $z \approx 4.35$~m, covering the rapidity of $-1.8 < \eta < 1.35$. 
It has 48 azimuthal sectors, where each sector contains six AstroPix layer slots (4 instrumented) interleaved with six Pb/ScFi calorimeter layers, plus six additional Pb/ScFi layers in the outer radial region (total 12 Pb/ScFi layers). 

Two readout systems are combined in the BIC. The AstroPix imaging layers use AstroPix silicon pixel sensors\cite{Kim:2025yyh} with 500 $\mu$m pitch, where each pixel records a hit time and ToT (Time-over-Threshold) to estimate the deposited energy, providing high-granularity 5D tracking of shower development. 
%It is a single-sample readout, with one measurement row per pixel firing. 
%The Pb/ScFi calorimeter layers consist of matrix of Pb (lead) glue and scintillating fibers that are read out at both longitudinal ends by SiPMs. 
The Pb/ScFi calorimeter consists of a lead/scintillating-fiber sampling structure read out at both longitudinal ends by SiPMs~\cite{Klest:2025pza}.
Each sector contains 60 SiPMs per side, with 5 SiPMs reading out each layer. The SiPM signals are digitized by the CALOROC-like digitization electronics as waveforms sampled at 25 ns intervals.
 %~\cite{HGCROC}. 
The SiPM analog signal is first digitized by the front-end electronics, producing a waveform representation of the pulse. Each waveform consists of multiple time samples (n samples) taken at fixed 25 ns intervals. In this readout scheme, every sample is recorded as a separate measurement row, while all samples belonging to the same pulse share a common \texttt{hit} index. Below, we discuss the detailed design of the AI architecture.

% \begin{figure}[ht]
%     \centering
%     \includegraphics[width=0.7\textwidth]{figures/cartoon.pdf}
%     \caption{Data processing workflow for (a) AstroPix and (b) Pb/ScFi in the BIC, from particle interaction to AI-ready data.}
%     \label{fig:cartoon}
% \end{figure}

\subsection{Schema Specialization}
\subsubsection{AI-ready data array}
% Applying the \textcolor{brown}{\texttt{measurements}} schema to the BIC requires four adaptations:
% \begin{enumerate}
%     \item The \texttt{detector} field is implemented as \texttt{subsystem}, with index 0 for AstroPix and 1 for Pb/ScFi. Because the two subsystems have different spatial resolutions, timing semantics, and signal physics, AI models should learn separate representations for each, for example by using \texttt{subsystem} as a learned embedding index.
%     \item The \texttt{sample} sub-index is zero for all AstroPix hits (single-sample readout). For Pb/ScFi, indices $0$ through $k-1$ label the $k$ successive 25~ns ADC waveform samples from the same SiPM channel. An additional time-of-arrival (TOA) entry, derived from the CALOROC fine-timing measurement, is appended at index $k$. The TOA row carries a sub-nanosecond arrival time in its \texttt{time} field but zero \texttt{deposit\_energy}, distinguishing it from the ADC samples.
%     \item The general \texttt{amplitude} field is stored as \texttt{deposit\_energy} in physical units of GeV.
%     \item The six-component position covariance matrix is simplified to three diagonal scalar uncertainties (\texttt{x\_unc}, \texttt{y\_unc}, \texttt{z\_unc}), since the BIC readout unit positions carry no significant off-diagonal correlations.
% \end{enumerate}

Applying the \textcolor{brown}{\texttt{measurements}} schema to the BIC requires four adaptations. First, the \texttt{detector} field is implemented as \texttt{subsystem}, with index 0 for AstroPix and 1 for Pb/ScFi. Because these two subsystems differ in spatial resolution, timing semantics, and signal physics, AI models should read separate representations for each, for example by treating \texttt{subsystem} as a learned embedding index. A detailed description of the data metadata structure is provided in the Appendix for reference.

Second, the \texttt{sample} sub-index is different for the two subsystems. It is mostly indexed zero for AstroPix hits, reflecting its dominant single-sample signal specification and readout scheme. 
For Pb/ScFi, indices $0$ through $k-1$ correspond to the $k$ successive 25~ns ADC waveform samples from the same SiPM channel, with each sample assigned a time based on the clock ($t_0$) as $t_i = t_0 + i \cdot 25$~ns.
An additional time-of-arrival (TOA) entry derived from the CALOROC fine-timing measurement is appended at index $k$.
This TOA row carries an arrival time in its \texttt{time} field at the TOA threshold amplitude, distinguishing it from the ADC samples. 

Third, the general \texttt{amplitude} field is represented as \texttt{deposit\_energy}, calibrated to physical units of GeV. For AstroPix, it corresponds to the calibrated energy converted from the time over threshold (ToT) measurement. For ScFi, it corresponds to the energy deposited in the scintillating fibers, after accounting for light attenuation along the fiber and the SiPM photon detection efficiency, with optional additional smearing applied to model detector response effects. In addition, the TOA threshold is converted from ADC units to the same GeV unit calibration scale as the waveform samples, ensuring a consistent energy representation across all derived quantities. 

Finally, the six-component position covariance matrix is reduced to three diagonal scalar uncertainties (\texttt{x\_unc}, \texttt{y\_unc}, \texttt{z\_unc}), since the BIC readout unit positions do not exhibit significant off-diagonal correlations.

\begin{table}[ht]
\centering
\caption{Structure of the \textcolor{brown}{\texttt{measurements}} array for the BIC.}
\begin{tabular}{lll}
\toprule
\hline \hline
Field & Type & Description \\
\midrule
\hline
\texttt{event}                   & \texttt{uint64}  & Event index defined per simulation collision. \\
\texttt{subsystem}               & \texttt{uint16}  & 0 = AstroPix, 1 = Pb/ScFi. \\
\texttt{hit}                     & \texttt{uint16}  & A single channel response from AstroPix or Pb/ScFi waveform. \\
\texttt{sample}                  & \texttt{uint16}  & Usually 0 for AstroPix. 0--N for Pb/ScFi waveform sample. \\
\texttt{x, y, z}                 & \texttt{float32}$\times$3 & Position of the AstroPix hit or Pb/ScFi SiPM readout (mm). \\
\texttt{deposit\_energy}         & \texttt{float32} & Deposited energy (GeV) converted from ToT or waveform ADC. \\
\texttt{time}                    & \texttt{float32} & Readout time at the measurement (ns). \\
\texttt{x\_unc, y\_unc, z\_unc}  & \texttt{float32}$\times$3 & Diagonal position uncertainties (mm). \\
\texttt{deposit\_energy\_unc}    & \texttt{float32} & Deposited energy uncertainty from calibrated resolution (GeV). \\
\texttt{time\_unc}               & \texttt{float32} &  Uncertainty on the measurement time from DAQ(ns). \\
\hline \hline
\bottomrule
\end{tabular}
\label{table:bic_measurements}
\end{table}

The resulting BIC \textcolor{brown}{\texttt{measurements}} array is summarized in Table~\ref{table:bic_measurements}.
The interpretation of the shared fields differs between subsystems. For AstroPix, \texttt{x, y, z} are the center coordinates of the pixel, \texttt{deposit\_energy} is the calibrated energy deposited, and \texttt{time} is the time recorded. 

For Pb/ScFi, the measurement coordinates encode the readout-channel location rather than the particle energy-deposition point. 
The physical longitudinal hit position is not directly measured from a single-end signal and can only be inferred from timing and/or amplitude information from both fiber ends.
Accordingly, the Pb/ScFi \texttt{x, y, z} coordinates correspond to the positions of the SiPM readouts at the two ends of the fiber-bundle grid cells. 
The \texttt{x} and \texttt{y} coordinates are given by the geometric center of the grid cell, while \texttt{z} is set by the SiPM location at the positive or negative $z$ end (1762.5 or $-2587.5$ mm).
Because the Pb/ScFi region is coupled to light guides at both ends, the stored $z$ values appear only at these two discrete positions.
The actual particle interaction position along the fiber can be reconstructed from the time difference between the two ends, $z_{\rm hit} = (t_{+z} - t_{-z}) \cdot v_{\rm eff}/2$, where $t_{+z}$ and $t_{-z}$ are the signal arrival times at the $+z$ and $-z$ fiber ends, respectively, and $v_{\rm eff} = 160$~mm/ns is an effective light-propagation velocity that includes the optical path in the fiber and detector response effects. %(refractive index $n \approx 1.59$). 
The \texttt{deposit\_energy} is converted from ADC using the calibration constant independently for each 25~ns waveform sample, and \texttt{time} denotes the electronic waveform sampling time, including fiber propagation delay. 

The reconstructed hit uncertainties are defined in terms of the Cartesian position uncertainties $\sigma_x$, $\sigma_y$, and $\sigma_z$, the deposited energy uncertainty $\sigma_E$, and the timing uncertainty $\sigma_t$. Their definitions depend on the detector subsystem. %For AstroPix, the position uncertainties are taken to be $\sigma_x = \sigma_y = \frac{0.5~\mathrm{mm}}{\sqrt{12}} \approx 0.144~\mathrm{mm}$, while $\sigma_z$ is set to zero with no uncertainty.
For AstroPix, the pixel plane is perpendicular to the radial direction. 
The intrinsic in-plane resolutions are  $\sigma_\varphi = \sigma_z =  0.5~\mathrm{mm}/\sqrt{12} \approx 0.144~\mathrm{mm}$ , determined by the pixel pitch, while the radial resolution is $\sigma_r = 5.0~\mathrm{mm}/\sqrt{12} \approx 1.44~\mathrm{mm}$ based on the
active sensor thickness, assuming a uniform distribution over the pixel sensor. 
The corresponding Cartesian uncertainties are obtained by rotating $(\sigma_r,\sigma_\varphi)$ by the sector azimuthal angle $\varphi_{\rm sec}$:
\begin{align} 
\sigma_x = \sqrt{\sigma_r^2\cos^2\varphi_{\mathrm{sec}} + \sigma_\varphi^2\sin^2\varphi_{\mathrm{sec}}}, 
\qquad \sigma_y = \sqrt{\sigma_r^2\sin^2\varphi_{\mathrm{sec}} +\sigma_\varphi^2\cos^2\varphi_{\mathrm{sec}}}.
\end{align}
The deposited energy uncertainty is obtained from deposit energy ($E_{\mathrm{dep}}$) based on the beam-test data calibration and parameterized as \begin{equation} 
\sigma_E = \begin{cases} 0.0457\,E_{\mathrm{dep}}, & E_{\mathrm{dep}} > 1.22 × 10^{-4}~\mathrm{GeV}, \\
5.89 × 10^{-8}\,E_{\mathrm{dep}}^{-0.499}, & E_{\mathrm{dep}} \leq 1.22 × 10^{-4}~\mathrm{GeV}, \end{cases} 
\end{equation} 
with the additional requirement $\sigma_E \leq E_{\mathrm{dep}}$; otherwise, $\sigma_E$ is set to 1. 
The timing uncertainty is fixed at $\sigma_t = 10~\mathrm{ns}$ consistent with the AstroPix timing resolution. 

For Pb/ScFi, the transverse position uncertainties are first defined in the local radial--azimuthal frame. Let $\sigma_r$ denote half of the radial cell size and $\sigma_{\mathrm{tan}}$ half of the azimuthal cell size. 
The uncertainties in the global Cartesian frame are then obtained by rotation through the sector azimuthal angle $\varphi_{\mathrm{sec}}$: 
\begin{equation} 
\sigma_x = \sqrt{ \sigma_r^2 \cos^2\varphi_{\mathrm{sec}} + \sigma_{\mathrm{tan}}^2 \sin^2\varphi_{\mathrm{sec}} }, \qquad \sigma_y = \sqrt{ \sigma_r^2 \sin^2\varphi_{\mathrm{sec}} + \sigma_{\mathrm{tan}}^2 \cos^2\varphi_{\mathrm{sec}} }. 
\end{equation} 
Here, the radial cell size is 20~mm, while the azimuthal cell size varies from 22 to 31~mm from the inner to the outer radius. 
The longitudinal uncertainty is taken to be $\sigma_z = 0$ since the readout is assigned to the discrete SiPM positions at the two fiber ends. The energy and timing uncertainties are currently set to $\sigma_E = 0, \sigma_t = 0$ pending final calibration constants.
All the detector-specific meanings are documented in the metadata.

\subsubsection{Truth Labeling}

Applying the \textcolor{brown}{\texttt{labels}} array to the BIC simulation data requires a few adaptations. For calorimeter truth labeling, each shower secondary produced within the BIC is traced back to the contributing particle, defined as the stable generator-level particle that entered the BIC volume and initiated the corresponding shower. 
It represents the truth label of a measured hit, such as an AstroPix hit or a Pb/ScFi SiPM waveform, for which multiple samples share the same truth label. 
The deposited energies from all Geant4 steps associated with the same contributing particle are then summed within the readout unit, while their positions and times are combined using energy-weighted averages. 
In AstroPix, single-contribution hits dominate, while in the Pb/ScFi calorimeter, overlapping showers in $\phi$ and $z$ more frequently produce multiple contributions within the same SiPM readout. 
Nevertheless, retaining up to $N = 3$ contributions still captures over 99\% of all cases across both subsystems. The true parent--daughter relationships are also preserved through the external MC generator tables. 
The label structure is given in Table~\ref{table:bic_labels}.
%In the AstroPix, single-contribution hits dominate, while in the Pb/ScFi, shower development allows $N = 3$ still captures over 99\% of all contributions across both subsystems, as discussed in the next part. 

\begin{table}[ht]
\centering
\caption{Structure of the \textcolor{brown}{\texttt{labels}} array for the BIC (contributor fields repeated for $n \in \{0, 1, 2\}$).}
\begin{tabular}{lll}
\toprule
\hline \hline
Field & Type & Description \\
\midrule
\hline
\texttt{event}  & \texttt{uint64}  & Event index. \\
\texttt{subsystem}  & \texttt{uint16}  & Subsystem identifier. \\
\texttt{hit}  & \texttt{uint16}  & Hit index. \\
\texttt{sample}  & \texttt{uint16}  & Sample index. \\
\texttt{x, y, z}                    & \texttt{float32}$\times$3 & Energy-weighted truth position (mm). \\
\texttt{deposit\_energy}            & \texttt{float32} & Total truth energy deposit (GeV). \\
\texttt{time}                       & \texttt{float32} & Earliest truth interaction time (ns). \\
\texttt{n\_contributions}           & \texttt{uint16}  & Total contributing particles. \\
\texttt{con\{n\}\_particle}         & \texttt{uint16}  & Index into event-level truth record. \\
\texttt{con\{n\}\_x, y, z}         & \texttt{float32}$\times$3 & Truth interaction position (mm). \\
\texttt{con\{n\}\_deposit\_energy}  & \texttt{float32} & Deposited energy from contributor $n$ (GeV). \\
\texttt{con\{n\}\_time}             & \texttt{float32} & Truth interaction time (ns). \\
\texttt{con\{n\}\_px, py, pz}       & \texttt{float32}$\times$3 & Momentum at production (GeV/$c$). \\
\texttt{con\{n\}\_energy}           & \texttt{float32} & Energy at production (GeV). \\
\texttt{con\{n\}\_pid}              & \texttt{int32}   & PDG ID. \\
\texttt{con\{n\}\_parent\_pid}      & \texttt{int32}   & PDG ID of parent. \\
\texttt{con\{n\}\_parent\_energy}   & \texttt{float32} & Parent energy (GeV). \\
\hline \hline
\bottomrule
\end{tabular}
\label{table:bic_labels}
\end{table}

% A contributor particle fetching algorithm (\emph{HitProcessor}) is applied to both subsystems to ensure that labels reflect the correct incident particle. 
% For each contribution, the algorithm walks the particle chain backward until it finds the first particle that (i) has not yet entered the calorimeter volume and (ii) has generator status equal to~1 (i.e., a stable final-state particle from the event generator). 
The per-contribution kinematic quantities, including \texttt{con\{n\}\_pid}, \texttt{con\{n\}\_px, py, pz}, and \texttt{con\{n\}\_energy} are taken from the contributing particles' MC generator truth information.
The variables \texttt{con\{n\}\_x, y, z} denote the true interaction positions of the contributing particle on the pixel or individual fiber, and are therefore distinct from the measured \texttt{x}, \texttt{y}, and \texttt{z} values stored in \textcolor{brown}{\texttt{measurement}}.
 In \textcolor{brown}{\texttt{labels}}, the quantities \texttt{x}, \texttt{y}, \texttt{z}, and \texttt{E} are defined from all contributions in the hit using energy-weighted averages, while \texttt{t} records the earliest arrival time among all contributions associated with that measurement. 
 All detector-specific parameters are documented in the metadata.

For completeness, the BIC dataset also includes an \textcolor{brown}{\texttt{event\_labels}} array that stores event-level MC particle records in Table~\ref{table:label}. This array is the lookup table referenced by \texttt{con\{n\}\_particle} in \textcolor{brown}{\texttt{labels}}, enabling retrieval of full particle kinematics and decay chain reconstruction. Since this information can also be obtained from standard formats such as HepMC3~\cite{Buckley:2019xhk}, the \textcolor{brown}{\texttt{event\_labels}} array is not required by the general framework.

\begin{table}[ht]
\centering
\caption{Structure of the event-level truth record array for the BIC.}
\begin{tabular}{lll}
\toprule
\hline\hline
Field & Type & Description \\
\hline\midrule
\texttt{event}         & \texttt{uint64}  & Event index. \\
\texttt{particle}      & \texttt{uint16}  & True particle index within the event. \\
\texttt{Vx, Vy, Vz}   & \texttt{float32} & True production vertex (mm). \\
\texttt{px, py, pz}   & \texttt{float32} & True particle momentum at production (GeV/$c$). \\
\texttt{energy}        & \texttt{float32} & True particle energy at production (GeV). \\
\texttt{time}          & \texttt{float32} & Production time (ns). \\
\texttt{charge}        & \texttt{int8}    & Electric charge. \\
\texttt{pid}           & \texttt{int32}   & PDG ID. \\
\texttt{parentnumbers} & \texttt{uint8}   & Number of parent particles. \\
\texttt{parentpid}     & \texttt{int32}   & PDG ID of the parent. \\
\texttt{parentindex}   & \texttt{uint16}  & Index of the parent in this array. \\
\hline\hline
\bottomrule
\end{tabular}
\label{table:label}
\end{table}

\subsection{Data Preparation and Format}
We produced the BIC simulation dataset of neutral-current deep-inelastic scattering (NC-DIS) $e+p$ events at 10~GeV $\times$ 100~GeV with $Q^2_{\min} = 1-1000$~GeV$^2$ generated by \texttt{PYTHIA8}~\cite{Sjostrand:2014zea}. The events were passed to \texttt{GEANT4}~\cite{Agostinelli:2002hh} simulation of full ePIC detector (citation) within the \texttt{dd4hep}~\cite{Frank:2014zya} detector framework, and the resulting simulated data were then processed with \texttt{eicrecon}~\cite{eicrecon} for digitization and hit reconstruction.

% The data preparation pipeline converts output from the PYTHIA8\,--\,GEANT4\,--\,\texttt{eicrecon} simulation chain into the AI-ready format.
% %, as illustrated in Figure~\ref{fig:cartoon}. 
We obtained reconstructed AstroPix-layer readouts and sampled digitized waveforms for the Pb/ScFi calorimeter readouts. For the AstroPix, the truth association follows a direct chain. Each reconstructed hit is linked via Geant4 indices to simulated MC particles, each containing per-particle time and energy contributions as listed in Table.~\ref{table:bic_labels}. 
For the Pb/ScFi, the truth association follows a more complex digitization chain. 
Each digitized waveform corresponds to the summed response of a single Pb/ScFi SiPM channel, which can contain overlapping single pulses (1-to-$N$), where each single pulse originates from multiple Geant4 truth contributions associated with different particles (1-to-$N$).
%Each digitized waveform maps to a combined pulse, which maps to multiple individual pulses at each SiPM, each associated with Geant4 truth contributions aggregated by particle index.

The dataset is stored as structured NumPy arrays in compressed \texttt{.npz} format. Each file contains three arrays: \textcolor{brown}{\texttt{measurements}}, \textcolor{brown}{\texttt{labels}}, and \textcolor{brown}{\texttt{event\_labels}}. 
%The \sout{hierarchical} field structure is illustrated in Figure~\ref{fig:tree_structure}. 
An accompanying metadata file provides descriptions of every field, enabling interpretation by users and AI models without prior knowledge of detector physics. A representative entry from the \textcolor{brown}{\texttt{measurements}} metadata is shown below: 
\begin{verbatim} 
 - **subsystem:** `uint16` categorical index, dimensionless. Identifies the
detector subsystem: 0 = AstroPix silicon pixel layers, 1 = Pb/ScFi calorimeter. 
This is the BIC specialization of the general `detector` field. Use this field 
as a learned embedding index (analogous to a token-type ID in a transformer) or 
to route hits through subsystem-specific encoder branches, since the two 
subsystems have fundamentally different spatial resolutions, timing semantics, 
and signal physics.
\end{verbatim} 

% \begin{figure}[ht]
%     \centering
%     \includegraphics[width=0.47\textwidth]{figures/DIS-pythia8NCDIS_10x100_minQ2_1000_5_planA_label.png}
%     \caption{Tree structure of the AI-ready \texttt{.npz} dataset, showing the three arrays and their constituent fields.}
%     \label{fig:tree_structure}
% \end{figure}

Each \texttt{.npz} file stores 1k events ($\sim$80k hit samples total) with size of 6~MB compressed. 
This format is used to enable a direct loading into NumPy and straightforward integration with PyTorch \texttt{Dataset} classes. 
The fixed-width structured arrays allow batch-level random access without parsing overhead, and avoid the event-by-event iteration inside the tree-structure.

\subsection{Visualization}

\begin{figure}[ht]
\centering
\begin{minipage}[t]{0.48\linewidth}
\centering
\includegraphics[width=\linewidth]{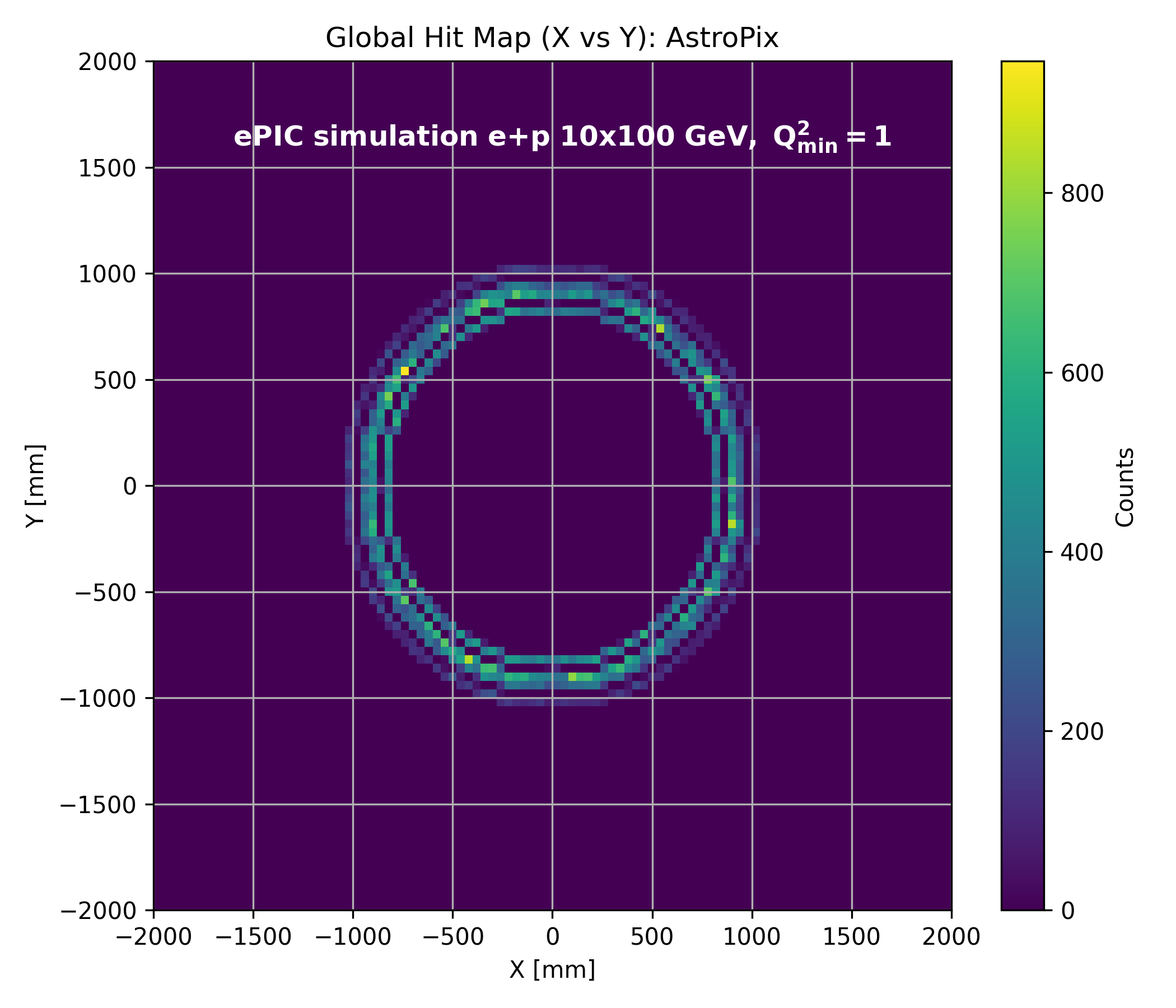}
\center{(a) AstroPix silicon pixel layers.}
\end{minipage}
\hfill
\begin{minipage}[t]{0.48\linewidth}
\centering
\includegraphics[width=\linewidth]{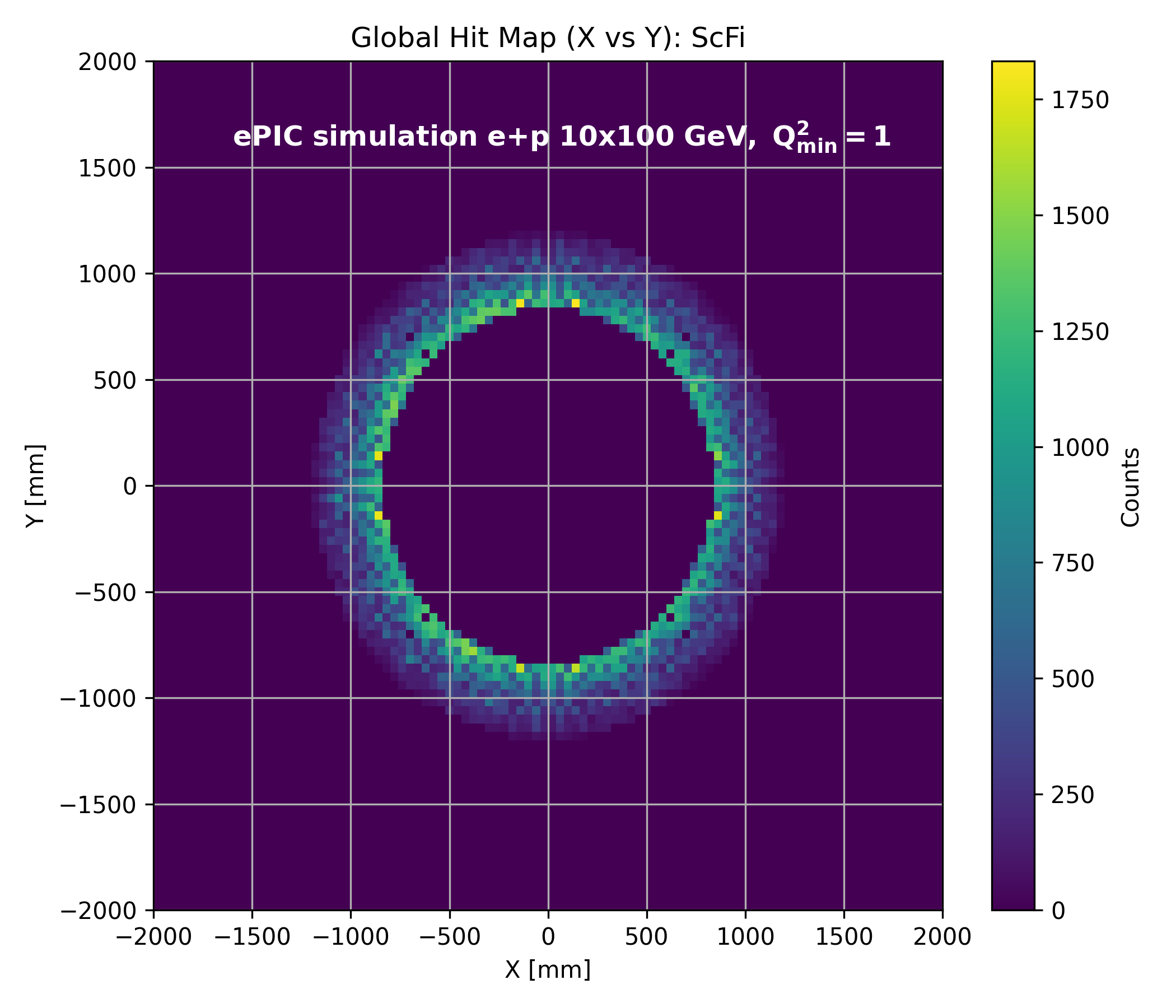}
\center{(b) Pb/ScFi calorimeter.}
\end{minipage}
\caption{Global transverse ($x$--$y$) hit map for the two BIC subsystems. The ring structure confirms that the stored positions represent physical readout unit locations, consistent with the BIC cylindrical geometry.}
\label{fig:bic_xy_map}
\end{figure}

For 5k NC-DIS events with $Q^2_{min}=1$, Fig.~\ref{fig:bic_xy_map} shows the transverse ($x$--$y$) hit distribution for both subsystems. The AstroPix hits populate the inner annulus at radii $r \approx 0.825$--$1.0$~m, while the Pb/ScFi SiPM readout units further occupy the outer annulus at $r \approx 0.825$--$1.2$~m. The ring structure and azimuthal uniformity across 48 sectors confirm that the stored coordinates correctly reflect the physical detector geometry.

%Figure~\ref{fig:bic_energy_corr} 
%shows the correlation between the AstroPix hit energy in \textcolor{brown}{\texttt{measurements}} and the truth energy deposit in \textcolor{brown}{\texttt{labels}}. The linear correspondence confirms correct measurement-to-label association and energy conversion, the slope is the sampling fraction of AstroPix layer with $s.f. = 0.00429453$.
% shows the correlation between the AstroPix hit energy in \textcolor{brown}{\texttt{measurements}} and the corresponding truth energy deposited in the AstroPix active layers as recorded in \textcolor{brown}{\texttt{labels}}. 
% The linear correspondence confirms correct measurement-to-label association and energy conversion. 
% % The slope reflects the effective sampling fraction of the AstroPix layer for the shower within the full calorimeter system, with $s.f. = 0.00429453$.
% The slope reflects the calibration factor used in the current reconstruction, where the AstroPix energy is defined as the particle deposit energy scaled by a fixed sampling fraction of $0.00429453$.

% \begin{figure}[ht]
% \centering
% \includegraphics[width=0.5\textwidth]{figures/pythia8NCDIS_10x100_minQ2_1_1/hits_AstroPix_EnergyCorrelation.png}
% \caption{Correlation between the AstroPix hit deposited energy in \textcolor{brown}{\texttt{measurements}} and the truth energy deposit in \textcolor{brown}{\texttt{labels}}. The linear trend is based on sampling constant = 0.00429453. }
% \label{fig:bic_energy_corr}
% \end{figure}

\begin{figure}[ht]
\centering
\begin{minipage}[t]{0.48\linewidth}
\centering
\includegraphics[width=\linewidth]{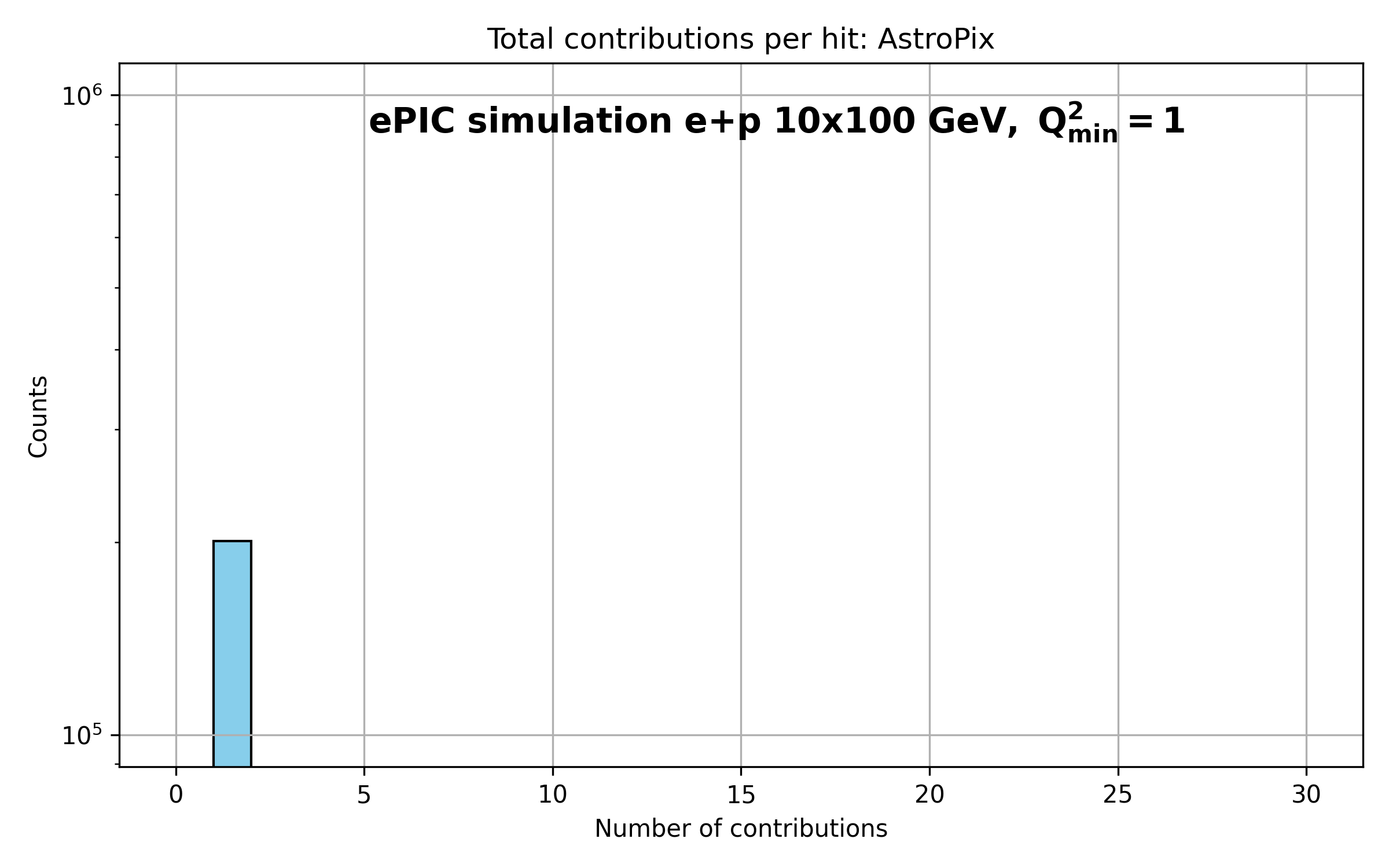}
\center{(a) AstroPix silicon pixel layers.}
\end{minipage}
\hfill
\begin{minipage}[t]{0.48\linewidth}
\centering
\includegraphics[width=\linewidth]{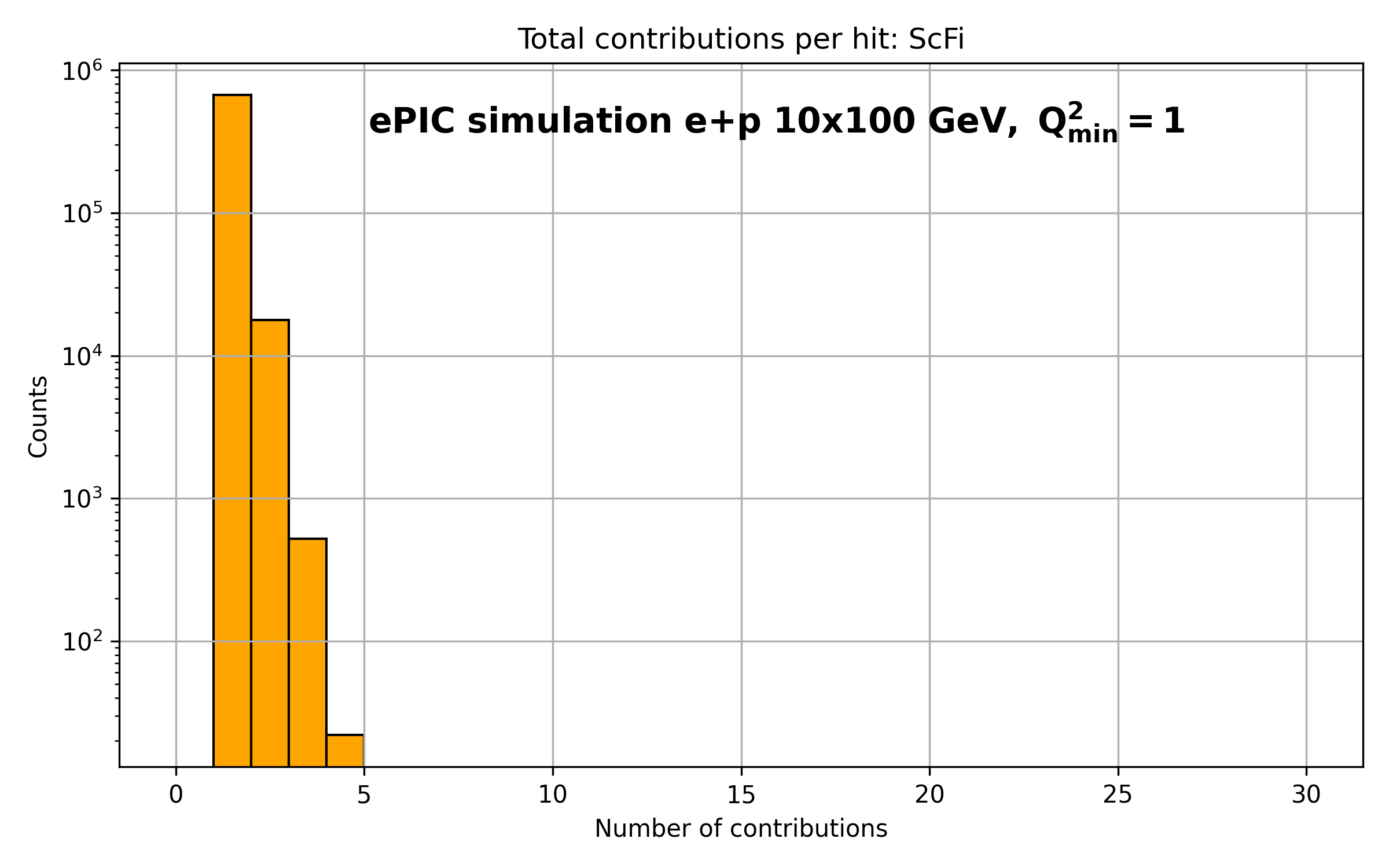}
\center{(b) Pb/ScFi layers.}
\end{minipage}
\caption{Distribution of the number of MC particle contributions per hit for the two BIC subsystems. Single-contribution hits dominate in both cases; the choice of $N = 3$ contribution slots covers over 99\% of all hits.}
\label{fig:bic_ncon}
\end{figure}

Figure~\ref{fig:bic_ncon} shows the distribution of contributions per hit. In the AstroPix, most hits receive a single particle, reflecting the tracker-like role of the pixel layers. In the Pb/ScFi, a larger fraction of multi-contribution hits arises from shower development in the lead absorber and the larger SiPM integration volume. With $N = 3$ contribution slots, the format covers over 99\% of all hit contributions in both subsystems.

\section{Summary and Outlook}
\label{sec:summary}

We have presented a detector-agnostic design for AI-ready data in nuclear and particle physics, motivated by the heterogeneous structure of QCD detector readout. The format is organized around two core arrays: \textcolor{brown}{\texttt{measurements}}, which stores detector responses in a unified spacetime-signal schema, and \textcolor{brown}{\texttt{labels}}, which keeps the corresponding measurement-level truth information from contributing particles. These arrays provide a compact and extensible representation that preserves physically relevant information while supporting scalable downstream AI applications.

With simulated data from the Barrel Imaging Calorimeter at ePIC detector, we showed that the general framework can accommodate a hybrid detector combining two different readout technologies, including AstroPix silicon pixel imaging layers and Pb/ScFi calorimetry, within a single schema through targeted specializations. 
The preliminary data visualizations confirm correct geometry and adequate coverage of contribution multiplicity with $N = 3$ retained contributors. Furthermore, this design supports both supervised learning and unsupervised learning, and serves as a point cloud representation for graph neural networks or transformer based foundation models.

Beyond its general utility for AI applications, the curated measurement array is also designed to be compatible with streaming data processing, aligning with the shift towards triggerless and streaming-style readout architectures at the LHC and EIC.
Its compact $(x,y,z,A,t)$ structure is advantageous for such environments because it provides a uniform, low-complexity representation of detector responses while explicitly retaining time as a continuous feature for ordering, association, and real-time reconstruction. 
Since the ePIC experiment is pursuing streaming-capable readout, it will be a good testing point for validating the representation is suitable for triggerless data acquisition workflows in the future.

\begin{acknowledgments}{The material is based upon work partially supported by the U.S. Department of Energy, Office of Science, Office of Nuclear Physics under the Contract No. DE-AC0206CH11357 and the American Science Cloud (AmSC) Data Providers Program (DaPP).} \end{acknowledgments}
\newpage

\appendix
\section{ePIC BIC Dataset Metadata}
\label{sec:metadata}

The dataset metadata provides natural-language descriptions of every field in each array, enabling correct interpretation by users and AI models without prior knowledge of detector physics. The metadata is stored as a structured markdown file accompanying each \texttt{.npz} data file. The following descriptions serve as the basis for the final metadata document.

\subsection{Metadata for the \texttt{measurements} Array}

\begin{itemize}

\item \textbf{\texttt{event}:} \texttt{uint64} integer index, dimensionless. A unique identifier for each physics event, corresponding to one simulated electron--proton collision. All rows from both AstroPix and Pb/ScFi subsystems belonging to the same collision share this index. This is a grouping key for batching, not a learnable feature. Rows are sorted by \texttt{event}. Not a learnable feature.

\item \textbf{\texttt{subsystem}:} \texttt{uint16} categorical index, dimensionless. Identifies the detector subsystem: 0 = AstroPix silicon pixel layers, 1 = Pb/ScFi calorimeter. This is the BIC specialization of the general \texttt{detector} field. Use this field as a learned embedding index (analogous to a token-type ID in a transformer) or to route hits through subsystem-specific encoder branches, since the two subsystems have fundamentally different spatial resolutions, timing semantics, and signal physics. Not a learnable feature.

\item \textbf{\texttt{hit}:} \texttt{uint16} integer index, dimensionless. Sequential index of each distinct readout signal within an event. For AstroPix, one hit corresponds to one pixel firing above threshold. For Pb/ScFi, one hit corresponds to one SiPM channel waveform; multiple waveform samples from the same channel share the same \texttt{hit} index and are distinguished by \texttt{sample}. Together, (\texttt{event}, \texttt{subsystem}, \texttt{hit}) uniquely identifies a readout signal.

\item \textbf{\texttt{sample}:} \texttt{uint16} integer index, dimensionless. Sub-index within a hit. For AstroPix, always 0. For Pb/ScFi, indexes successive 25~ns waveform samples from the same SiPM channel; rows sharing the same \texttt{hit} but with increasing \texttt{sample} form an ordered time sequence representing the digitized waveform shape. An additional TOA (Time of Arrival) entry is appended at sample index $n+1$.

\item \textbf{\texttt{x, y, z}:} \texttt{float32} $\times$ 3, spatial scalars, unit: mm. Cartesian coordinates of the readout unit in the laboratory frame, with the $z$-axis along the proton beam direction. For AstroPix, this is the center of the fired pixel within the silicon bulk. For Pb/ScFi, this is the position of the SiPM readout unit at the longitudinal fiber end, not the physical location of the particle interaction along the fiber. The SiPM readout $x,y$ location is defined at the center of the grid cell, and its $z$ position is at one of the two ScFi ends: $z = -2587.5$~mm or $1762.5$~mm. The true interaction $z_{\mathrm{hit}}$ must be reconstructed from the time difference between the two fiber ends ($t_{+z}$ and $t_{-z}$), as $z_{\mathrm{hit}} = (t_{+z} - t_{-z}) \cdot v_{\mathrm{eff}} / 2$, based on the effective velocity of $v_{\mathrm{eff}} = 160$~mm/ns. Coordinate convention: the nominal interaction point is at the origin $(0, 0, 0)$.

\item \textbf{\texttt{deposit\_energy}:} \texttt{float32} scalar, unit: GeV. The measured signal amplitude converted to deposited energy. For AstroPix, it is derived from the charge collected by the pixel and corresponds to calibrated ToT (time over threshold). For Pb/ScFi, it is energy attenuated along the fiber to the ends of the detector, which is converted independently for each 25~ns waveform sample from either the 14-bit ADC value (dynamic range 0.1~MeV to 10~GeV in true energy) or the electric photon count (1700~e.p.\ is 0.1~GeV in deposited energy not corrected for sampling fraction). The energy spans approximately five orders of magnitude ($\sim10^{-5}$ to $\sim10$~GeV); logarithmic scaling is recommended before training. The truth counterpart is \texttt{deposit\_energy} in \texttt{labels}.

\item \textbf{\texttt{time}:} \texttt{float32} scalar, unit: ns. The electronic readout time associated with the measurement. For AstroPix, this is the time-of-arrival of the charge signal. For Pb/ScFi, this is the waveform sampling time $t_{\mathrm{sample}} = t_{\mathrm{phase}} + k \cdot 25$~ns, which includes both the initial particle interaction time and the light propagation delay from the interaction point to the SiPM; it is \emph{not} the physical hit time on the fiber. For Pb/ScFi rows sharing the same \texttt{hit}, the \texttt{time} values form an ordered sequence at 25~ns intervals, representing the waveform shape. The truth counterpart is \texttt{time} in \texttt{labels}.

\item \textbf{\texttt{x\_unc, y\_unc, z\_unc}:} \texttt{float32} $\times$ 3 scalars, unit: mm. One-sigma diagonal position uncertainties on the \texttt{x}, \texttt{y}, \texttt{z} columns respectively (simplified from the general \texttt{pos\_cov} covariance matrix to three diagonal scalars because the BIC readout positions carry no significant off-diagonal correlations). For AstroPix: $\sigma_z = 0.5\;\mathrm{mm}/\sqrt{12}$ in the global Cartesian coordinate system, and $\sigma_x$, $\sigma_y$ are obtained by rotating the uncertainties from the active-layer thickness of 5~mm in the radial direction and 0.5~mm in the azimuthal direction by the azimuthal angle. For Pb/ScFi: $\sigma_x$ and $\sigma_y$ are set by the SiPM readout unit size and rotated at its sector azimuthal angle. Each SiPM readout unit is 20~mm in the radial direction and 22--31~mm in the azimuthal direction depending on radius. The uncertainty is taken as the RMS of the corresponding unit size divided by $\sqrt{12}$. $\sigma_z = 0$~mm as $z$ is the position at the end of the SiPM sector.

\item \textbf{\texttt{deposit\_energy\_unc}:} \texttt{float32} scalar, unit: GeV. One-sigma uncertainty on the \texttt{deposit\_energy} column. For AstroPix, parameterized from bench test data: $\sigma_E \approx 0.0457 \cdot E_{\mathrm{dep}}$ for $E_{\mathrm{dep}} > 1.22 \times 10^{-4}$~GeV, otherwise $\sigma_E = 5.89 \times 10^{-8} \cdot E_{\mathrm{dep}}^{-0.499}$~GeV. For Pb/ScFi, pending calibration and is currently zero.

\item \textbf{\texttt{time\_unc}:} \texttt{float32} scalar, unit: ns. One-sigma uncertainty on the \texttt{time} column. For AstroPix, 10~ns based on chip timing resolution. For Pb/ScFi, these values are currently set to zero pending calibration; here, zero denotes ``not yet assigned'' rather than a physical value.

\end{itemize}

\subsection{Metadata for the \texttt{labels} Array}

\begin{itemize}

\item \textbf{\texttt{event}:} \texttt{uint64} integer index, dimensionless. Matches the \texttt{event} index in \texttt{measurements}. This is a join key: rows with the same \texttt{event} in \texttt{measurements} and \texttt{labels} belong to the same collision. Not a learnable feature.

\item \textbf{\texttt{subsystem}:} \texttt{uint16} categorical index, dimensionless. Matches the \texttt{subsystem} field in \texttt{measurements}: 0 = AstroPix, 1 = Pb/ScFi. This is a join key used together with \texttt{event}, \texttt{hit}, and \texttt{sample} for row-level association between the two arrays. Not a learnable feature.

\item \textbf{\texttt{hit}:} \texttt{uint16} integer index, dimensionless. Matches the \texttt{hit} index in \texttt{measurements}. Together with (\texttt{event}, \texttt{subsystem}, \texttt{sample}), this provides the unique key for joining each label row to its corresponding measurement row.

\item \textbf{\texttt{sample}:} \texttt{uint16} integer index, dimensionless. Matches the \texttt{sample} index in \texttt{measurements}. For AstroPix, always 0. For Pb/ScFi, indexes the waveform sample within a hit, ensuring truth information is available at the individual waveform sample level.

\item \textbf{\texttt{x, y, z}:} \texttt{float32} $\times$ 3 spatial scalars, unit: mm. Energy-weighted average truth position over all particle contributions to this measurement. This provides a single aggregate truth reference point for the interaction location. For AstroPix, this is close to the Geant4 simulated hit position within the pixel. For Pb/ScFi, $z$ is the physical interaction position along the fiber, and $x$ and $y$ are the cell readout location, in contrast to the SiPM readout position stored in \texttt{measurements}. Coordinate system: same Cartesian laboratory frame as \texttt{measurements}, with the $z$-axis along the proton beam direction and origin at the nominal interaction point.

\item \textbf{\texttt{deposit\_energy}:} \texttt{float32} scalar, unit: GeV. Total truth energy deposited in this measurement, summed over all contributing particles. This is the truth counterpart to the \texttt{deposit\_energy} column in \texttt{measurements}, and the sum of all \texttt{con\{n\}\_deposit\_energy} values equals this field up to rounding from contributions beyond MAX\_CONTRIB.

\item \textbf{\texttt{time}:} \texttt{float32} scalar, unit: ns. The earliest arriving truth interaction time among all particle contributions to this measurement. This is the truth counterpart to the \texttt{time} column in \texttt{measurements}. For Pb/ScFi, this is the physical hit time on the fiber, not the electronic waveform sampling time.

\item \textbf{\texttt{n\_contributions}:} \texttt{uint16} integer count, dimensionless. Total number of distinct particles that deposited energy in this measurement. May exceed MAX\_CONTRIB = 3; when it does, only the top 3 contributors by deposited energy are stored in the \texttt{con\{n\}} slots. This count encodes the degree of signal overlap: a value of 1 indicates a clean single-particle hit, while higher values indicate combined contribution from multiple particles. Particles are the smallest interaction carriers recorded by the detector. Particles produced at the event vertex are defined as primary particles, while those produced from the decay of a particle are defined as secondary particles. Here, a contribution particle refers to the incident particle entering the detector system, not to particles generated later by interactions inside the detector material, such as shower electrons.

\item \textbf{\texttt{con\{n\}\_particle}:} \texttt{uint16} integer index, dimensionless, where $n \in \{0, 1, 2\}$. Unique event-level index of the $n$-th contributing particle, serving as a foreign key into the \texttt{event\_label} array. Use this to retrieve the full kinematics of the contributing particle via the \texttt{particle} field in \texttt{event\_label}. Slots beyond total number of \texttt{n\_contributions} contain the sentinel value 0 and should be masked. For example, \texttt{n\_contributions} = 2, only slots 0 and 1 contain valid data. The later slots, such as slot 2, are filled with 0 as a placeholder and should be ignored or masked. Contributors are ranked by deposited energy (slot 0 is the largest contributor).

\item \textbf{\texttt{con\{n\}\_x, y, z}:} \texttt{float32} $\times$ 3 spatial scalars, unit: mm. True interaction position at silicon pixel or fiber-level granularity. For AstroPix, \texttt{con\{n\}\_x}, \texttt{y}, and \texttt{z} are not energy-weighted. For ScFi, \texttt{con\{n\}\_x}, \texttt{con\{n\}\_y}, and \texttt{con\{n\}\_z} give the energy-weighted interaction position on the same fiber from all shower particles produced by the $n$-th contributing particle in the detector. Notably, this is the physical interaction position along the fiber for $x$, $y$, and $z$, in contrast to the SiPM readout position stored in \texttt{measurements}. The $x$, $y$ may vary within the cell with offsets up to $\sim$10~mm from cell center. The $z$ varies over hundreds of mm along the barrel. Sentinel: $(0, 0, 0)$ for empty slots.

\item \textbf{\texttt{con\{n\}\_deposit\_energy}:} \texttt{float32} scalar, unit: GeV. Energy deposited by the $n$-th contributing particle in a single pixel or a single fiber. For ScFi only: summed over all shower particles by energy-weighted within the fiber. The sum over all valid contributors of this value would approximate the aggregate \texttt{deposit\_energy} field in this array. Contributors are sorted by this field in descending order. Sentinel: 0 for empty slots.

\item \textbf{\texttt{con\{n\}\_time}:} \texttt{float32} scalar, unit: ns. True energy-weighted interaction time of the $n$-th contributing particle. For Pb/ScFi, this is the physical hit time at the fiber, not the electronic waveform sampling time in \texttt{measurements}. Sentinel: 0 for empty slots.

\item \textbf{\texttt{con\{n\}\_px, py, pz}:} \texttt{float32} $\times$ 3 momentum scalars, unit: GeV/$c$. Three-momentum of the $n$-th contributing particle at its production vertex. In the BIC, a primary particle fetching algorithm traces each contribution back through the shower to the incident particle; these values are fetched from that traced particle, not the immediate shower secondary. Sentinel: $(0, 0, 0)$ for empty slots.

\item \textbf{\texttt{con\{n\}\_energy}:} \texttt{float32} scalar, unit: GeV. Total energy of the $n$-th contributing particle at its production vertex. Together with \texttt{con\{n\}\_px, py, pz}, this completes the particle four-vector. As with momentum, the value corresponds to the traced incident particle after primary particle fetching. Sentinel: 0 for empty slots.

\item \textbf{\texttt{con\{n\}\_pid}:} \texttt{int32} categorical identifier, dimensionless. Unique particle data group Monte Carlo numbering identifier\footnote{\url{https://pdg.lbl.gov/2007/reviews/montecarlorpp.pdf}}. Common codes: 11 = $e^-$, $-11$ = $e^+$, 22 = $\gamma$, 211 = $\pi^+$, $-211$ = $\pi^-$, 111 = $\pi^0$, 2212 = $p$. In the BIC, this is the PDG ID of the traced incident particle after primary particle fetching. Sentinel: 0 for empty slots (PDG ID 0 is unassigned and serves as a safe invalid marker).

\item \textbf{\texttt{con\{n\}\_parent\_pid}:} \texttt{int32} categorical identifier, dimensionless. PDG ID of the immediate parent of the $n$-th contribution particle. Used to identify the physical origin of it (e.g., a photon from $\pi^0$ decays to 2 photons) and to distinguish prompt particles from decay products. Sentinel: 0 when it is a primary particle from the event vertex and there is no parent.

\item \textbf{\texttt{con\{n\}\_parent\_energy}:} \texttt{float32} scalar, unit: GeV. Vertex-level energy of the immediate parent of the $n$-th contributor. Combined with \texttt{con\{n\}\_parent\_pid}, this enables downstream models to assess the energy scale of the parent particle. Sentinel: 0 when it is a primary particle from the event vertex and there is no parent.

\end{itemize}

\subsection{Metadata for the \texttt{event\_label} Array}

\begin{itemize}

\item \textbf{\texttt{event}:} \texttt{uint64} integer index, dimensionless. Same event index as in \texttt{measurements} and \texttt{labels}, identifying the collision to which this particle belongs. This is a join key: use it together with \texttt{particle} to look up a specific MC particle from a specific event.

\item \textbf{\texttt{particle}:} \texttt{uint16} integer index, dimensionless. Sequential index of the particle within the event. This is the primary lookup key referenced by \texttt{con\{n\}\_particle} in \texttt{labels}: given a contribution index, select the row in \texttt{event\_label} with matching \texttt{event} and \texttt{particle} to retrieve the full particle record.

\item \textbf{\texttt{Vx, Vy, Vz}:} \texttt{float32} $\times$ 3 spatial scalars, unit: mm. Production vertex of the particle in the Cartesian laboratory frame (same coordinate system as \texttt{x, y, z} in \texttt{measurements}). Values near $(0, 0, 0)$ identify primary collision particles produced at the interaction point; non-zero values indicate decay products or secondary hadronic interaction products created within the detector material. The displacement from the origin encodes the decay length or interaction depth.

\item \textbf{\texttt{px, py, pz}:} \texttt{float32} $\times$ 3 momentum scalars, unit: GeV/$c$. Three-momentum of the particle at its production vertex. Together with \texttt{Energy}, these four values form the particle's relativistic four-momentum $(E, p_x, p_y, p_z)$, from which invariant mass, rapidity, and transverse momentum can be derived.

\item \textbf{\texttt{energy}:} \texttt{float32} scalar, unit: GeV. Total energy of the particle at its production vertex. Combined with \texttt{px, py, pz}, this completes the four-vector. The invariant mass satisfies $m^2 = E^2 - (p_x^2 + p_y^2 + p_z^2)$.

\item \textbf{\texttt{time}:} \texttt{float32} scalar, unit: ns. Production time of the particle. For primary collision particles this is near zero; for decay products or secondary interaction products, a non-zero value indicates the particle was produced at a later time. This distinguishes prompt particles from delayed decays within the detector.

\item \textbf{\texttt{charge}:} \texttt{int8} integer scalar, unit: elementary charge ($e$). Electric charge of the particle. Typical values: $-1$ ($e^-$), $+1$ ($e^+$, $p$, $\pi^+$), 0 ($\gamma$, $n$, $\pi^0$). This enables models to distinguish charged particles (which leave tracks in the AstroPix pixel layers) from neutral particles (which do not).

\item \textbf{\texttt{pid}:} \texttt{int32} categorical identifier, dimensionless. PDG Monte Carlo numbering scheme identifier, uniquely specifying the particle species\footnote{\url{https://pdg.lbl.gov/2007/reviews/montecarlorpp.pdf}}. Common codes: 11 = $e^-$, $-11$ = $e^+$, 22 = $\gamma$, 211 = $\pi^+$, $-211$ = $\pi^-$, 111 = $\pi^0$, 2212 = $p$, 2112 = $n$. This is the primary label for particle identification tasks and is essential for interpreting decay chains.

\item \textbf{\texttt{parentnumbers}:} \texttt{uint8} integer count, dimensionless. Number of parent particles in the MC event record. A value of 0 identifies a primary collision product (no parent); a value of 1 indicates a standard decay or secondary interaction product. This allows downstream models to classify particles as primary or secondary without traversing the full ancestry chain.

\item \textbf{\texttt{parentpid}:} \texttt{int32} categorical identifier, dimensionless. PDG ID of the immediate parent particle. Combined with \texttt{parentindex}, this enables reconstruction of decay chains (e.g., identifying photons from $\pi^0$ decays to $\gamma+\gamma$ via \texttt{parentpid} = 111) without requiring the full HepMC3 event record. Set to 0 for primary particles with no parent.

\item \textbf{\texttt{parentindex}:} \texttt{uint16} integer index, dimensionless. Index of the immediate parent particle in this \texttt{event\_label} array (same \texttt{event}). This is a self-referencing foreign key: following \texttt{parentindex} recursively reconstructs the full decay chain back to the primary collision particle. The parent index is stored rather than daughter indices because a single decay typically produces multiple daughters. Set to 0 for primary particles.

\end{itemize}

\bibliography{references}

\end{document}